\label{key}
\documentclass[regular]{jfm}
\usepackage{graphicx}
\usepackage{newtxtext}
\usepackage{newtxmath}
\usepackage{natbib}
\usepackage{hyperref}
\usepackage{bm}
\hypersetup{
    colorlinks = true,
    urlcolor   = blue,
    citecolor  = blue,
}

\newcommand{\RomanNumeralCaps}[1]
\linenumbers

\newcommand\Ra{\mbox{\textit{Ra}}}

\newcommand\Prn{\mbox{\textit{Pr}}}

\newcommand\Am{\mbox{\textit{a}}}

\title{Flow structure transition in thermal vibrational convection}

\author{Xi-Li. Guo\aff{1},
  Jian-Zhao Wu\aff{1}\corresp{\email{jianzhao\_wu@shu.edu.cn}},
  Bo-Fu Wang\aff{1,2},
  Quan Zhou\aff{1,2},
  \and Kai Leong Chong\aff{1,2}\corresp{\email{klchong@shu.edu.cn}}}

\affiliation{\aff{1}Shanghai Institute of Applied Mathematics and Mechanics, School of Mechanics and Engineering Science, and Shanghai Key Laboratory of Mechanics in Energy Engineering, Shanghai University, Shanghai 200072, China \\
\aff{2} Shanghai Institute of Aircraft Mechanics and Control, Zhangwu Road, Shanghai 20092, China}

\begin{document}
%\graphicspath{{/Users/xili/Desktop/JFM/figure/}}
%\graphicspath{{/Users/figure/}}
\graphicspath{{./figure/}}
\maketitle

\begin{abstract}
This study investigates the effect of vibration on the flow structure transitions in thermal vibrational convection (TVC) systems, which occur when a fluid layer with a temperature gradient is excited by vibration. Direct numerical simulations of TVC in a two-dimensional enclosed square box were performed over a range of dimensionless vibration amplitudes $0.001 \le a \le 0.3$ and angular frequencies $10^{2} \le \omega \le 10^{7}$, with a fixed Prandtl number of 4.38. The flow visualisation shows the transition behaviour of flow structure upon the varying frequency, characterising three distinct regimes, which are the periodic-circulation regime, columnar regime and columnar-broken regime. Different statistical properties are distinguished from the temperature and velocity fluctuations at the boundary layer and mid-height. Upon transition into the columnar regime, columnar thermal coherent structures are formed, in contrast to the periodic oscillating circulation. These columns are contributed by merging of thermal plumes near the boundary layer, and the resultant thermal updrafts remain at almost fixed lateral position, leading to a decrease in fluctuations. We further find that the critical point of this transition can be described nicely by the vibrational Rayleigh number $\Ra_\mathrm{vib}$. As the frequency continues to increase, entering the so-called columnar-broken regime, the columnar structures are broken, and eventually the flow state becomes a large-scale circulation, characterised by a sudden increase in fluctuations.  Finally, a phase diagram is constructed to summarise the flow structure transition over a wide range of vibration amplitude and frequency parameters.

\end{abstract}

\begin{keywords}
Authors should not enter keywords on the manuscript, as these must be chosen by the author during the online submission process and will then be added during the typesetting process (see \href{https://www.cambridge.org/core/journals/journal-of-fluid-mechanics/information/list-of-keywords}{Keyword PDF} for the full list).  Other classifications will be added at the same time.
\end{keywords}

{\bf MSC Codes }  {\it(Optional)} Please enter your MSC Codes here

\section{Introduction} \label{sec:intro}

Fluid flow can be driven by various methods, such as buoyancy-driven \citep{Lohse2009RMP,Lohse2010ARFM,Xia2013TAML}, surface-tension-driven \citep{Levich1969ARFM,Kawamura2012JHT}, electromagnetic-driven \citep{Schussler2012PIAU,Macdonald2017APJ}, and vibration-driven \citep{Gershuni1998book,Lappa2009book} which is essential for many practical applications, such as heat transfer and mixing. While buoyancy-driven convection is a common way to induce fluid flow, in certain extreme conditions such as microgravity, it becomes challenging to generate sufficient flow due to the negligible effect of gravity. However, alternative methods such as vibration can become significant for heat transport, and vibration can be possibly caused by many ways in microgravity such as spacecraft movement,  astronaut activity \citep{Bannister1973apollo} or even manually imposed. Vibration can result in a supplementary fluid motion, making it an attractive way to control fluid flow and heat transfer in microgravity. In this context, thermal vibrational convection (TVC) arises when vibration is applied to a fluid system with a temperature gradient. The resulting body force causes relative fluid motion and increases the threshold for convective instability. Therefore, it is important to study the underlying flow physics of TVC as a potential way to drive fluid flow and heat transfer in microgravity. The present work focuses on investigating the effect of vibration on the structural transitions of the flow in TVC systems.
 
Many previous studies have focused on introducing vibration to the classical model of thermal convection, such as Rayleigh-B\'enard (RB) convection, to understand how boundary layer disruption affects heat transport. Vibration can either enhance or suppress convective heat transport depending on the mutual direction of the vibration axis and the temperature gradient \citep{Savino1998CF,Cisse2004IJHMT}. \citet{Wang2020SCIADV} applied horizontal vibrations to RB convection, and found that high frequency vibration can significantly enhance the heat transport. This enhancement is achieved by vibration-induced boundary layer (BL) destabilisation, which breaks up the limitation of laminar BL for heat transport. In contrast, when vertical vibration is applied to the turbulent RB convection, the convective heat transport is suppressed \citep{Wu2022JFM} by inducing the anti-gravity effect that stabilizes thermal BL and enhances thermal limitation. In addition to studying global quantities,  \cite{Wu2021POF} distinguish vibration-generated oscillatory flows and fluctuating fields in oscillatory RB convection with periodic lateral boundary conditions by using the phase decomposition method. It is shown that vibration not only induces strong shear forces that breaks thermal BL and triggers massive plume emissions, but also generates an oscillatory flow that enhances the intensity of the turbulent fluctuating flows in bulk zone.

The effect of vibration on vertical convection (VC) has also been studied. Vertical convection can be considered as RB convection with an extreme inclination angle of $90^{\circ} $ \citep{Shishkina2016JFM,Zwirner2018JFM,Zhang2021JFM,Wang2021JFM}, which indeed be encountered in some practical situations. A typical flow feature for VC  systems is the stably-stratified bulk region. \cite{XQGuo2022POF} studied numerically the effect of vertical vibration on VC systems. They revealed a critical vibration frequency $\omega^{*}$ beyond which heat transport enhancement is found. Besides the heat transport properties, the influence of vibration on the existing flow structure is also worthy to be studied. In the pure VC, the hot plumes move upwards along the heating wall, the cold plumes move downwards along the cooling wall, and the flow is self-organized to form the large-scale circulation (LSC). However, upon vibraiton, \cite{Bouarab2019POF} and \cite{Mokhtari2020PRF} showed that the LSC in the VC systems can be reversed when the angle between the vibration direction and gravity falls within two critical vibration angles.

As mentioned above, there have been many studies considering the joint action from both the buoyancy and vibration. With the vibration as an only driving force, there are also extensive studies especially in the field of aerospace. In spacecraft experiment, the vibration frequency can reach up to 100 Hz \citep{Marin2018MST,Dong2019npj} and a series of long-duration microgravity vibration experiments have been performed using spacecraft. Field experiments  on TVC in microgravity can be traced back to that on the Apollo spacecraft \citep{Grodzka1971apollo,Bannister1973apollo}. Their experiments revealed that vibrations caused by the vibrating motion of the spacecraft and the astronauts significantly increased the heat transfer compared to the case with pure conduction \citep{Grodzka1975Science}. Using the ALICE-2 instrument on the MIR station, \cite{Garrabos2007PRE} studied the effect of linear harmonic vibration on temperature propagation with a point-like heat source in near-critical fluid. Using experimental equipment on the International Space Station (ISS), \cite{Braibanti2019EPJE} studied the response of binary mixtures to the onboard g-jitter and vibrational forcing, respectively, where the density difference is introduced by thermal and compositional variations. Their study suggests that g-jitter on the ISS actually has a limited effect on thermal diffusion, and that externally imposed constant frequency and amplitude are the main factors affecting diffusive processes. A number of experiments \citep{Mialdun2008PRL,Shevtsova2010Acta,Shevtsova2010JFM} have been carried out in the microgravity environment emulated by the parabolic flight of the aircraft. It has been demonstrated that vibrations can induce convection in a non-uniformly heated fluid. This opens up a possible avenue for driving convection in microgravity environments. 

The morphological change of flow structure in TVC systems due to the onset of convective instability is also a key issue. Linear stability analysis based on mean flow dynamical equations have been performed to identify the relevant parameter characterizing the intensity of the vibrational excitation, which is the vibrational Rayleigh number $\Ra_\mathrm{vib}$ \citep{Gershuni1970JAMM,Gershuni1979}. \citet{Gershuni1982FD} examined the change of flow structure of the average vibrational convection under microgravity. When $\Ra_\mathrm{vib}$ is less than the critical value $\Ra_\mathrm{vib}^{*}$ ($\Ra_\mathrm{vib}^{*} / Pr \approx 1.5 \times 10^{4}$), there is steady four-vortex structure. As $\Ra_\mathrm{vib}$ increases, the four-vortex structure becomes an unstable mode and the structure transitions to a stable three-vortex structure. \citet{Crewdson2021IJT} performed a systematic parametric study of the of the mean velocity field that displaced a very regular columnar array of convective structures when the $\Ra_\mathrm{vib} \geq 5 \times 10^{7}$. In a numerical study of vibration over a wide range of frequencies and amplitudes, \citet{Hirata2001JFM} observed synchronous, subharmonic, and non-periodic responses of the flow to vibrational excitations. The non-periodic response may be related to the transition between flow states. \citet{Shevtsova2010JFM} performed the investigation of the effect of vibration on flow pattern bifurcation in a low-gravity environment using both two-dimensional 2D numerical modelling based on the averaging equations and three-dimensional direct numerical simulation (DNS) of TVC.

At present, the unified constitutive law of heat transport in TVC has been revealed in our previous work \citep{wu2022unifying}, however, the study on the morphological transition of flow structure in the turbulent state of TVC with microgravity is still lacking. To gain insight into this, a two-dimensional (2D) numerical study is carried out in a square enclosed domain with imposed temperature difference. Our aim is to study the transition of flow structure, especially in the unsteady and even turbulent state, and the associated statistical properties. The structure of this paper is organised as follows: In \S\ref{sec:level2}, the governing equations of TVC in microgravity are given, and the numerical methods and parameters are described. In \S\ref{sec:level3}, the flow structure transition in TVC is observed from the visualization of instantaneous and mean flow field. In \S\ref{sec:level4}, a quantitative analysis of statistical properties is given to characterise this transition and the transition phase is obtained. Finally, a brief conclusion is given in \S\ref{sec:level7}.

\section{\label{sec:level2}Numerical methods}\label{sec:Numerical methods}
 \begin{figure}
 	\centerline{\includegraphics[width=0.50\textwidth]{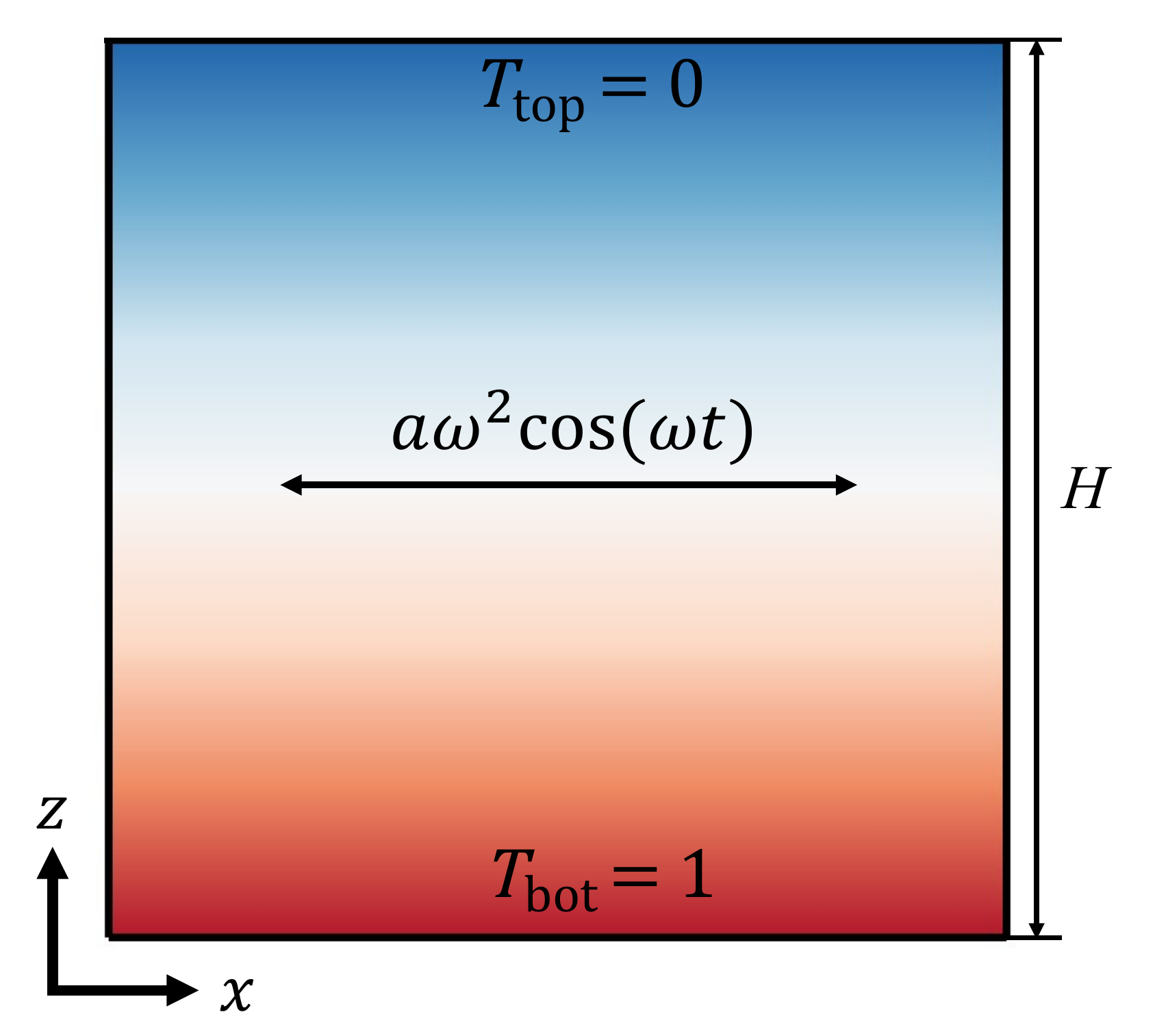}}  % Images in 100% size
 	\caption{Sketches of the 2D convection cell with the coordinate systems. The bottom conducting plate is heated with temperature $T_\mathrm{bot}$, while the top conducting plate is cooled with the temperature $T_\mathrm{top}$. The sidewalls are adiabatic. Vibration is applied to the cell in the horizontal direction. In the reference associated to vibration, an additional acceleration induced by vibration is introduced, i.e., $a \omega^{2} \cos(\omega t)\bm{e}_x$ in dimensionless form. }
 	\label{fig:model}
 \end{figure}

We carry out a series of DNS to study the vibration-driven thermal turbulence in a 2D square domain, as illustrated in figure~\ref{fig:model}. The external harmonic vibration is applied to the convection cell in the horizontal direction, which is perpendicular to the direction of temperature gradient. The normalisation is done by the scale of cell height $H$, the viscous diffusion time $H^{2} / \nu$, the temperature difference $\Delta$ between the top and bottom plates. The incompressible Oberbeck--Boussinesq equations for the TVC are given by \citep{Gershuni1998book,Shevtsova2010JFM}
\begin{equation} \label{eq:continuity}
	\nabla \cdot \pmb{u} = 0 ,
\end{equation}
\begin{equation} \label{eq:momentum}
	\frac{\partial \pmb{u}}{\partial t} +\pmb{u} \cdot \nabla \pmb{u} = - \nabla p + \nabla^2 \pmb{u} - \Am \omega^2 \cos(\omega t) T \pmb{e}_{x},
\end{equation}
\begin{equation} \label{eq:energy}
	\frac{\partial T}{\partial t} + \pmb{u} \cdot \nabla T = \Prn^{-1} \nabla^{2} T ,
\end{equation}
where, $\pmb{u}=(u, w)$, $T$ and $p$ are the dimensionless flow velocity, temperature and kinematic pressure, respectively, and $\pmb{e}_{x}$ is the unit vector along the horizontal direction. In equations \eqref{eq:continuity}--\eqref{eq:energy}, there are three control parameters, \textit{i.e.}, the Prandtl number $\Prn$, the dimensionless vibration amplitude $\Am$ and the dimensionless angular frequency $\omega$:
\begin{equation}
	\Prn = \frac{\nu}{\kappa}, \quad \Am = \frac{\alpha \Delta A}{H}, \quad \omega = \frac{\Omega H^{2}}{\nu}, 
\end{equation}
where $A$ is the pulsating displacement, $\Omega$ the angular frequency, and $\alpha$, $\nu$, $\kappa$ are the coefficients of thermal expansion, kinematic viscosity, and thermal diffusivity. Regarding the velocity boundary conditions, all solid boundaries are assumed to be impermeable and no-slip.  For temperature, an adiabatic condition is maintained at the sidewalls, while a constant temperature is applied to the top and bottom plates, \textit{i.e.}, $T_\mathrm{top} = 0$ and $T_\mathrm{bot} = 1$, respectively. 

In previous studies, the vibrational Rayleigh number $\Ra_\mathrm{vib}$, which is derived using the averaging approach under the condition of small amplitude and high frequency, has often been used to describe the averaged vibrational effects on convective flows. It is defined as:
\begin{equation}\label{eq:ravib}
	\Ra_\mathrm{vib} = \frac{(A \Omega \alpha \Delta H)^{2}}{2 \nu \kappa}.
\end{equation}

Numerical simulations are performed using the in-house finite-difference code, which has been well validated in our previous studies \citep{Wang2020SCIADV,Wu2022JFM}. A second-order finite-difference method is used for the spatial discretization in a staggered grid. The third order Runge-Kutta method combined with the second order Crank-Nicholson scheme is used for time integration. The pressure Poisson equation is solved using the fast-Fourier-transform (FFT) method. More details on the numerical approach can be found in \citet{zhang2019JH}. In all runs, we fixed $\Prn = 4.38$, which corresponds to the working fluid of water. The dimensionless amplitude ranges from $\Am = 0.001$ to $\Am = 0.3$ and the dimensionless frequency ranges from $\omega = 10^2$ to $\omega = 10^7$. The grid spacing increases with increasing amplitude $\Am$ and frequency $\omega$. For instance, the grid size is increased from $512 \times 512$ for $\omega = 10^2$ to $2048 \times 2048$ for $\omega = 10^7$ at fixed $a=0.1$, and from $1024 \times 1024$ for $\Am = 0.001$ to $2048 \times 2048$ for $\Am = 0.3$ at fixed $\omega=10^7$. The mesh is refined in the near-wall region to ensure accurate resolution of the near-wall dynamics within the oscillating Stokes layer. For simulation with high frequencies, the Stokes layer is resolved with at least $16$ meshes.  For the time integration, we maintained the time step $\Delta_{t} \leq \tau_\omega / 60$, where $ \tau_\omega = 2 \pi / \omega$, such that the convective flows during each vibration period are resolved with at least $60$ time steps. Moreover, the mesh and time step are well chosen following the resolution requirements of \citet{shishkina2010njp}.

\section{\label{sec:level3}Transition of flow structure} \label{Transitional phase of flow structure}

\begin{figure}
	\centerline{\includegraphics[width=1.0\textwidth]{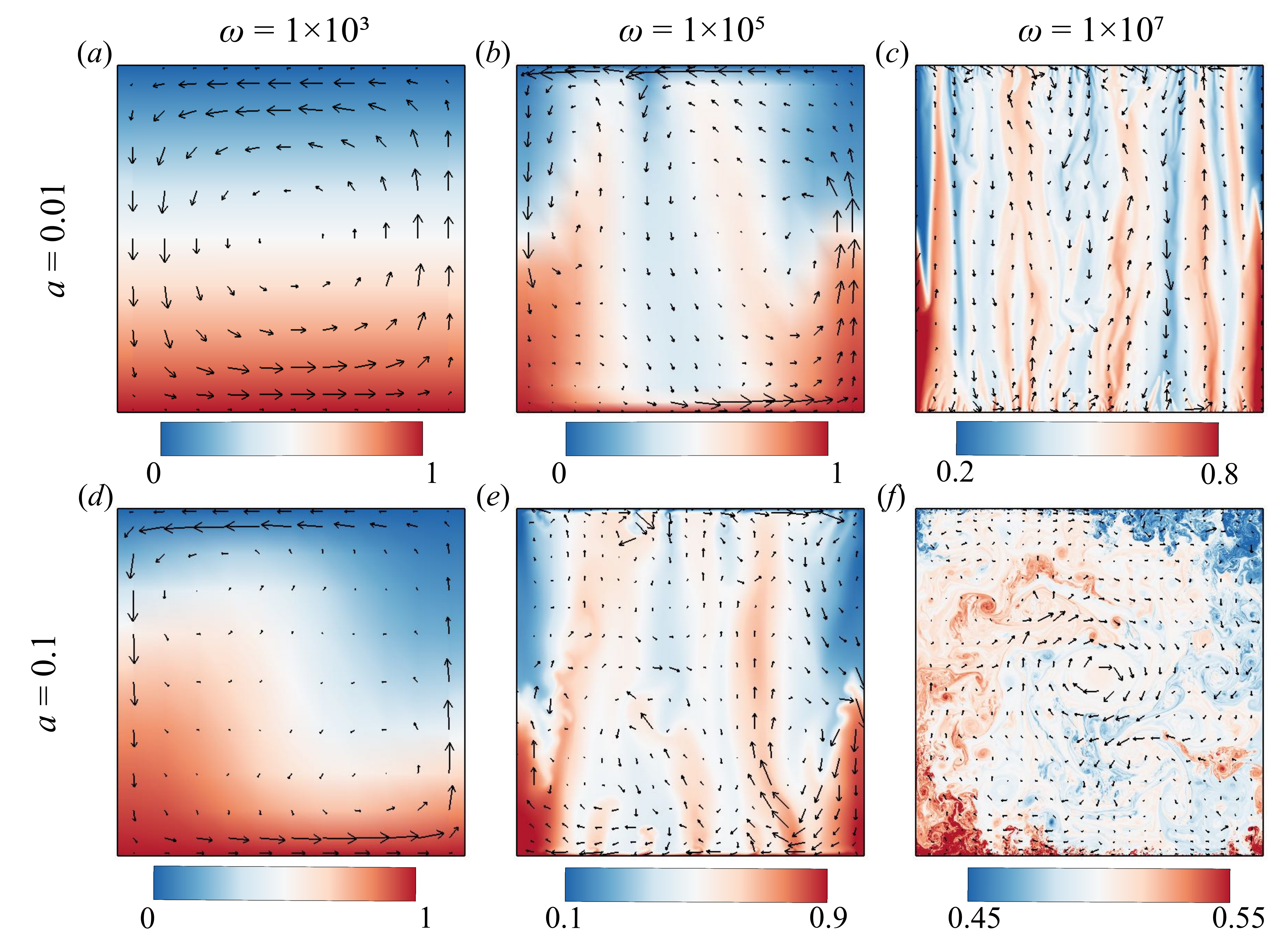}}	% Images in 100% size
	\caption{Typical snapshots of the instantaneous temperature (color) and velocity (vectors) fields for two different $a$, which are $a=0.01$ ($a$-$c$) and $a=0.1$ ($d$-$f$), and for three different $\omega$, which are $\omega=10^{3}$ ($a$,$d$), $\omega=10^{5}$ ($b$,$e$) and $\omega=10^{7}$ ($c$,$f$). The magnitude of the velocity is represented by the length of the arrows in non-dimensional units and the temperature is represented by the color.}
	\label{fig:inst}
\end{figure}

\begin{figure}
	\centerline{\includegraphics[width=1.0\textwidth]{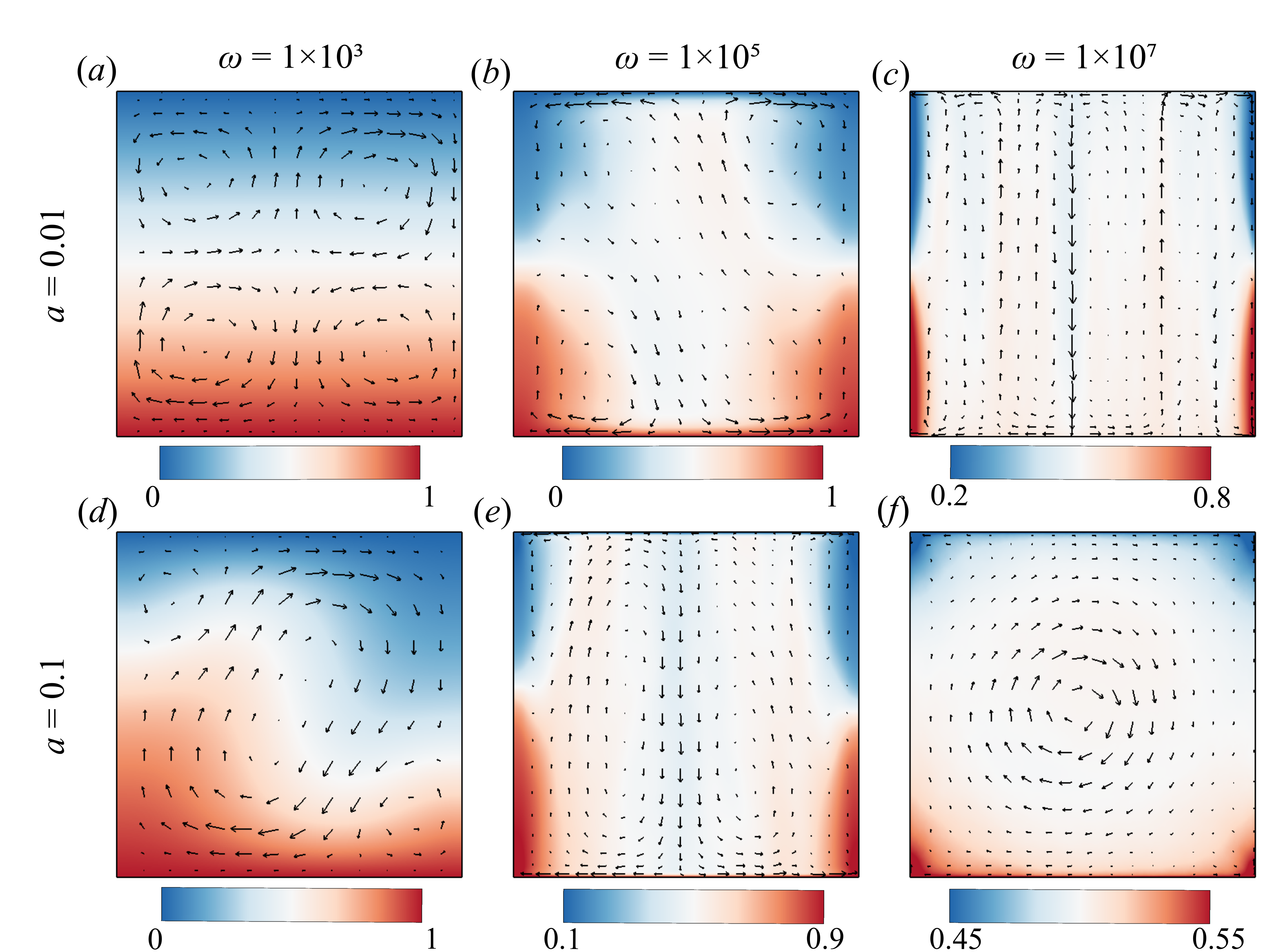}}	% Images in 100% size
	\caption{Typical snapshots of the time-average temperature (color) and velocity (vectors) fields for two different $a$, which are $a=0.01$ ($a$-$c$) and $a=0.1$ ($d$-$f$), and for three different $\omega$, which are $\omega=10^{3}$ ($a$,$d$), $\omega=10^{5}$ ($b$,$e$) and $\omega=10^{7}$ ($c$,$f$). The magnitude of the velocity is represented by the length of the arrows in non-dimensional units and the temperature is represented by the color.}
	\label{fig:mean}
\end{figure}

We analyse the flow structures through the visualization of the temperature and velocity fields. Figure~\ref{fig:inst} shows the snapshots of the instantaneous temperature and velocity fields for different vibration amplitudes ($a=0.01,0.1$) and frequencies ($\omega = 10^{3}, 10^{5}, 10^{7}$). As shown in figure~\ref{fig:inst}$(a)$, when the frequency is small ($\omega = 10^{3}$ at $a=0.01$), a nearly stable temperature distribution is established in the bulk. When the frequency reaches $\omega = 10^{5}$, the morphology of the flow changes significantly as shown in figure~\ref{fig:inst}($b$). The vibration destabilises the flow such that the flow state deviates from that of pure conduction, eventually the temperature field changes from a nearly linear temperature distribution to the one with a stack of columnar thermal plume. As the vibration frequency is increased further to $\omega = 10^{7}$, more fragmented hot (cold) plumes are ejected from the thermal boundary layer which subsequently merge into the columnar thermal updrafts (downdrafts). 

Apart from the dependence on frequency, the effect of vibration amplitude on flow structure has also been shown in figure~\ref{fig:inst}($d$-$f$). As the amplitude increases to $a=0.1$ from $a=0.01$, the flow field becomes more chaotic as revealed by the more fragmented and irregular temperature structure. The difference caused by increasing amplitude can even be more pronounced for $\omega = 10^{7}$. The columnar thermal structure observed in the case $a=0.01$ has been broken down as $a$ is increased to $0.1$, and one observes the transition to a flywheel structure as shown in figure~\ref{fig:inst}($f$) (see movie 3 in the supplementary movies), where the hot plume moves upwards on one side and the cold plume moves downwards on the other side.

To gain a deeper understanding on the flow structure, we examine the time-averaged fields of temperature and velocity as shown in figure~\ref{fig:mean}. At small $\omega$, the four-roll structure for $a=0.01$ (figure~\ref{fig:mean}$a$) and a three-roll structure for $a=0.1$ (figure~\ref{fig:mean}$d$) are established in the flow, which is consistent with the previous observations of TVC in a microgravity environment \citep{Mialdun2008PRL,Shevtsova2010JFM}. The observed four-roll flow pattern in the flow field is a result of the superposition of two counter-rotating circulations (see movie 1 in the supplementary movies), which are induced by the back-and-forth vibration. In figure~\ref{fig:mean}$(b,c,e)$, the clear footprint of the columnar thermal plume can also be seen in the time-averaged field. It signifies that the location of these heat transport channels are almost stable over time.  The flow structure then evolves into the LSC shown in figure~\ref{fig:mean}$(f)$.

%Fig.~\ref{fig:inst} shows transition from a state with almost linear temperature profile to columnar thermal structure, it resembles the case of quasi-2D convection. With increasing frequency $\omega$, we observe the stark transition to fly wheel structure, where hot plumes move upward on one side and the cold plumes move downward on the other side. To make sure the formation of large-scale circulation, we look at the time-averaged field Fig.~\ref{fig:mean}. From there, we see that the columnar structures are quite stable where the averaged field still show their footprint.

%4------------------------------------------------------------------------------------------------------------------------------------------

\section {\label{sec:level4} Statistical properties in different flow states} 
\subsection{Mean and standard deviation temperature profiles}

\begin{figure}
	\centerline{\includegraphics[width=1.0\textwidth]{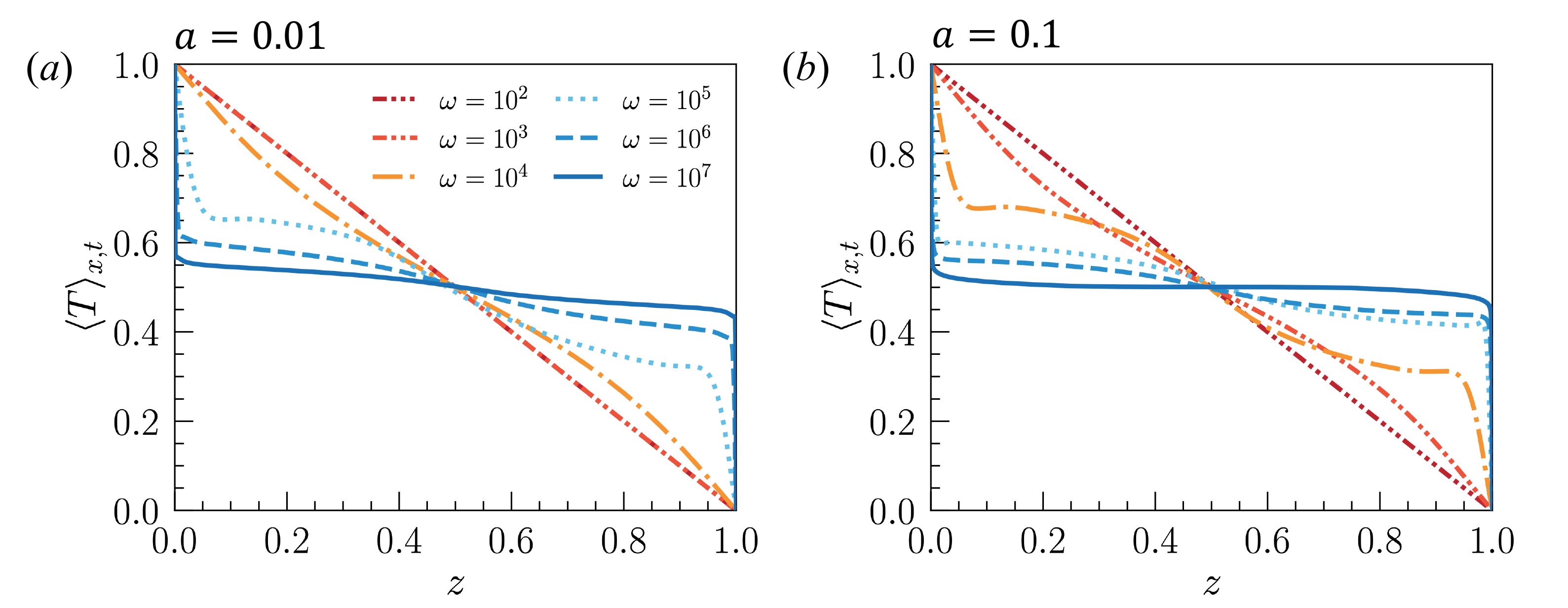}}% Here is how to import pdf art
	\caption{Vertical profiles of mean temperature for various $\omega$ at ($a$) $a=0.01$  and ($b$) $a=0.1$ .}
	\label{fig:temppro}
\end{figure}

\begin{figure}
	\centerline{\includegraphics[width=1.0\textwidth]{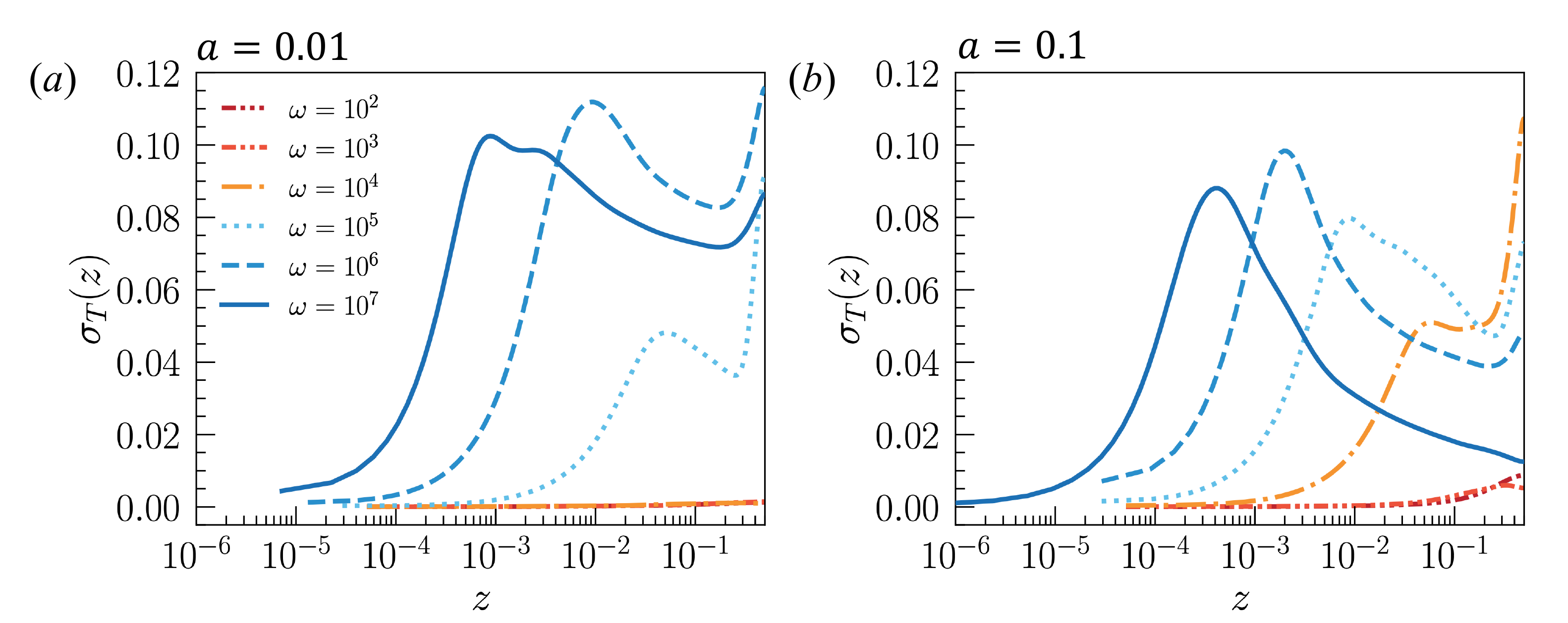}}% Here is how to import EPS art
	\caption{Vertical profiles of mean temperature SD $\sigma_{T}(z)$ for various $\omega$ at ($a$) $a=0.01$ and ($b$) $a=0.1$.}
	\label{fig:temprms} 
\end{figure}

As mentioned above, the global flow structure undergoes two kinds of transitions with the increasing control parameters $\omega$ at sufficiently large amplitude $a$. The first flow transition is characterised by a change from periodic circulating flow to a convective flow with columnar thermal structures, while the second transition is characterised by the emergence of flywheel structure. In this subsection, we quantify local statistics that correspond to the change of global flow structure.  

Figure~\ref{fig:temppro} depicts the temperature profiles at amplitudes $a=0.01$ and $a=0.1$ for various frequencies $\omega$. For small $\omega$, a linear distribution of temperature profiles is observed, which confirms the stable temperature distribution found in figure~\ref{fig:inst}$(a,d)$ and figure~\ref{fig:mean}$(a,d)$. As $\omega$ increases, the thermal BLs become distinctly visible near the top and bottom plates. Within the thermal BL, the temperature decreases (increases) sharply from normalised temperature $T_\mathrm{bot}=1$ ($T_\mathrm{top}=0$) in proximity to the bottom (top) plate to the bulk temperature. At the largest explored frequency $\omega$ of $a=0.1$, the temperature structure with thermal shortcut in the bulk is observed, reminiscent of that in RB convection \citep{Lohse2009RMP,Xia2013TAML}. Interestingly, at certain intermediate frequencies, the temperature change can be non-monotonic with an inversion layer existing at the junction between the bulk and the boundaries. This phenomenon is also observed in VC systems due to the presence of stable stratification in the bulk \citep{Ravi1994JFM,Wang2021JFM} and in the similarity solution of natural convection BL equations \citep{Henkes1989IJHMT}.

Next, the profiles of standard deviation (SD) for temperature $\sigma_{T}(z)$ are examined, which quantifies the temporal fluctuation of the temperature field, defined as follows:
\begin{equation}
	\sigma_{T}(z) = \Big \langle \sqrt{\langle T^{2} \rangle_{t} - \langle T \rangle_{t}^{2}} \Big \rangle_{x},
\end{equation}
where $\langle \cdot \rangle_{t}$ and $\langle \cdot \rangle_{x}$ are the time average and spatial average in horizontal direction, respectively. The vertical profile of temperature fluctuations, $\sigma_T$, as a function of the vertical distance $z$ away from the bottom plate drawn up to the mid-height is shown in figure ~\ref{fig:temprms}. As the frequency increases to sufficiently large values ($\omega=10^5$), the location of the edge of the thermal BL becomes apparent, as defined based on the position of the maximum temperature fluctuations \citep{Belmonte1994PRE}. With an increasing frequency, the thermal BL becomes thinner due to the increasing shear effect of vibration. Additionally, a secondary peak at the mid-height is observed. The flow field shown in figure~\ref{fig:mean} suggests that this secondary peak is related to the presence of the recirculating flow. Visualised by movie 2 in the supplementary movies, the merge of hot and cold columnar plumes leads to the intense mixing near the mid-height, which contribute to the significantly enhanced temperature fluctuation.

%4.2------------------------------------------------------------------------------------------------------------------------------------------
\subsection {Profiles of velocity standard deviation} %4.2

\begin{figure}
	\centerline{\includegraphics[width=1.0\textwidth]{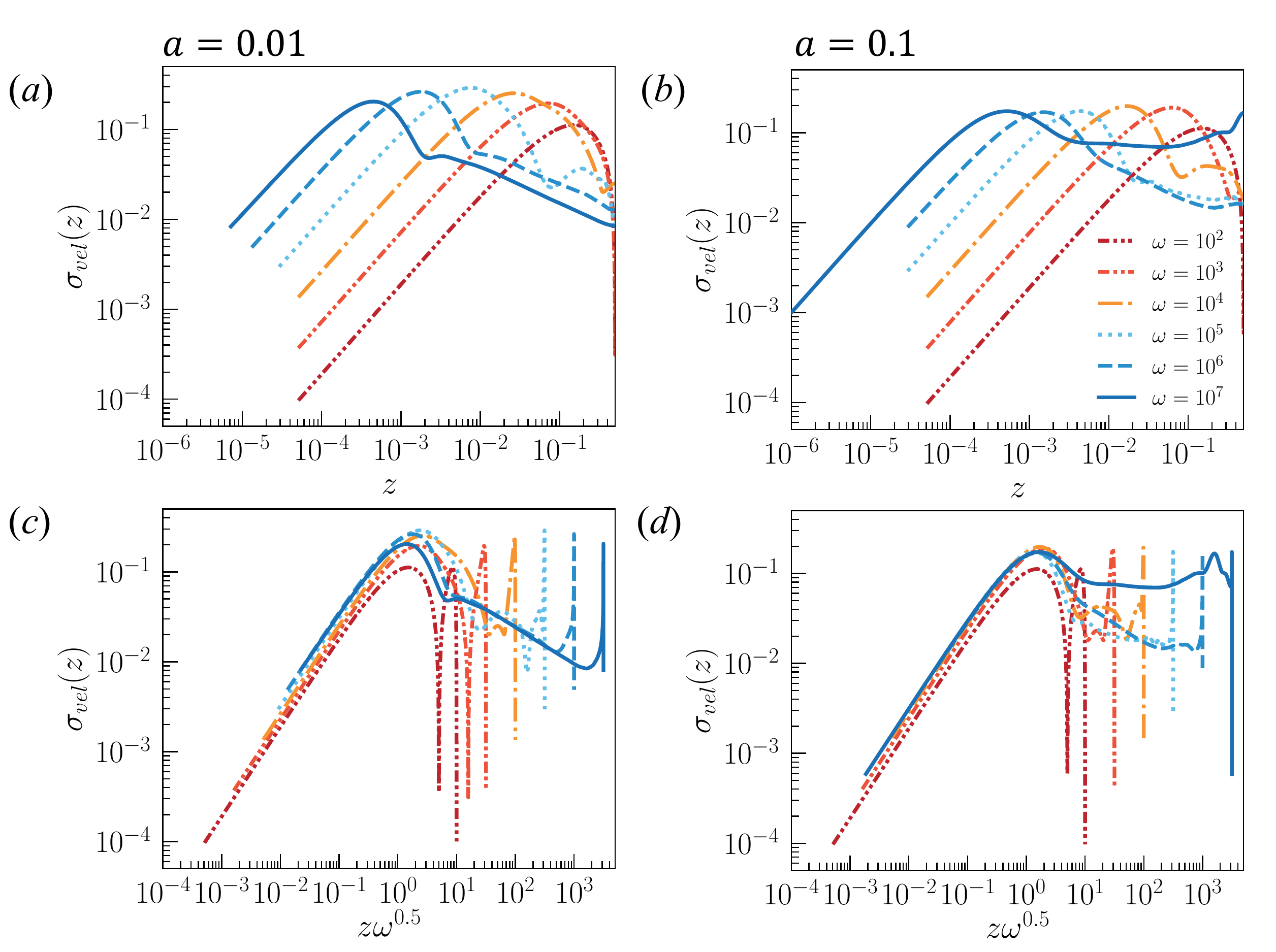}}% Here is how to import EPS art
	\caption{Mean velocity SD $\sigma_{\mathrm{vel}}(z)$ as a function of the normalized vertical distance $z$ ($a$-$b$) and $z \omega^{0.5}$ ($c$-$d$) under different vibration amplitude with $a=0.01$ ($a$,$c$) and $a=0.1$ ($b$,$d$).}
	\label{fig:urms_combo} 
\end{figure}

%     \begin{figure}
	%	\centerline{\includegraphics[width=13cm]{uRMS.pdf}}% Here is how to import EPS art
	%	\caption{Velocity RMS, as a function of the normalised vertical distance $z/H$.}
	%	\label{fig:urms} 
	%	\end{figure}

%     \begin{figure}
	%	\centerline{\includegraphics[width=13cm]{uRMS_normalize.pdf}}% Here is how to import EPS art
	%	\caption{Velocity RMS,as a function of the normalised vertical distance $z\omega^{0.5}$.}
	%	\label{fig:urms_nor} 
	%	\end{figure}

Besides the temperature profiles, we examine the velocity fluctuation using the modulus of the SD of all velocity components $\sigma_{\mathrm{vel}}$, which is calculated as 
\begin{equation}
	\sigma_\mathrm{vel}(z)=\sqrt{\sigma_u^2(z)+\sigma_w^2(z)},
\end{equation}
where $\sigma_u(z)$ and $\sigma_w(z)$ are the SDs of the horizontal and vertical velocities normalized by the pulsating velocity, respectively, with the following definitions:
\begin{equation}
	\sigma_u(z)=\frac{1}{a \omega}\left\langle\sqrt{\left\langle u^2\right\rangle_t-\langle u\rangle_t^2}\right\rangle_x,  \quad  \sigma_w(z)=\frac{1}{a \omega}\left\langle\sqrt{\left\langle w^2\right\rangle_t-\langle w\rangle_t^2}\right\rangle_x.
\end{equation}
In figure~\ref{fig:urms_combo},  we show the $\sigma_{\bm{u}}(z)$ profiles at two different amplitudes, namely, $a = 0.01$ (figure~\ref{fig:urms_combo}$a$,$c$) and $a = 0.1$(figure~\ref{fig:urms_combo}$b$,$d$). As $\omega$ increases, the velocity BL, defined by the height of the peak, becomes thinner. It is noteworthy that the magnitude of the peak and the shape of profiles near the peak look similar for different frequencies $\omega$ (figure~\ref{fig:urms_combo}$a$,$b$). To further confirm the universal properties existing near the BL, we adopt the normalisation using the thickness of the Stokes layer. As the disturbance generated by the oscillating plate travels as a transverse wave in the region of the fluid adjacent to the wall, and the penetration depth of this wave defines the Stokes layer, which is defined as $\delta_{\omega} = \sqrt{2\nu / \Omega} $ (namely $\delta_{\omega}/H = \sqrt{2/ \omega} $). We then plot the $\sigma_{\mathrm{vel}}(z)$ as a function of the normalised vertical distance $z\omega^{0.5}$ in figure~\ref{fig:urms_combo}($c$-$d$). The results show that the $\omega$-dependent peaks of $\sigma_{\mathrm{vel}}(z)$ all collapse onto the same curve for $z\omega^{0.5}\leq1$. As the height progressively reaches the bulk region, the velocity fluctuations decrease (cases with columnar flow structure) due to the diminishment of the vibrating effect travelled by the walls. However, once the broken of columnar flow takes place and flywheel structure emerges, the fluctuation $\sigma_{\mathrm{vel}}(z)$ in the bulk increases significantly.

\begin{figure}
	\centerline{\includegraphics[width=1.1\textwidth]{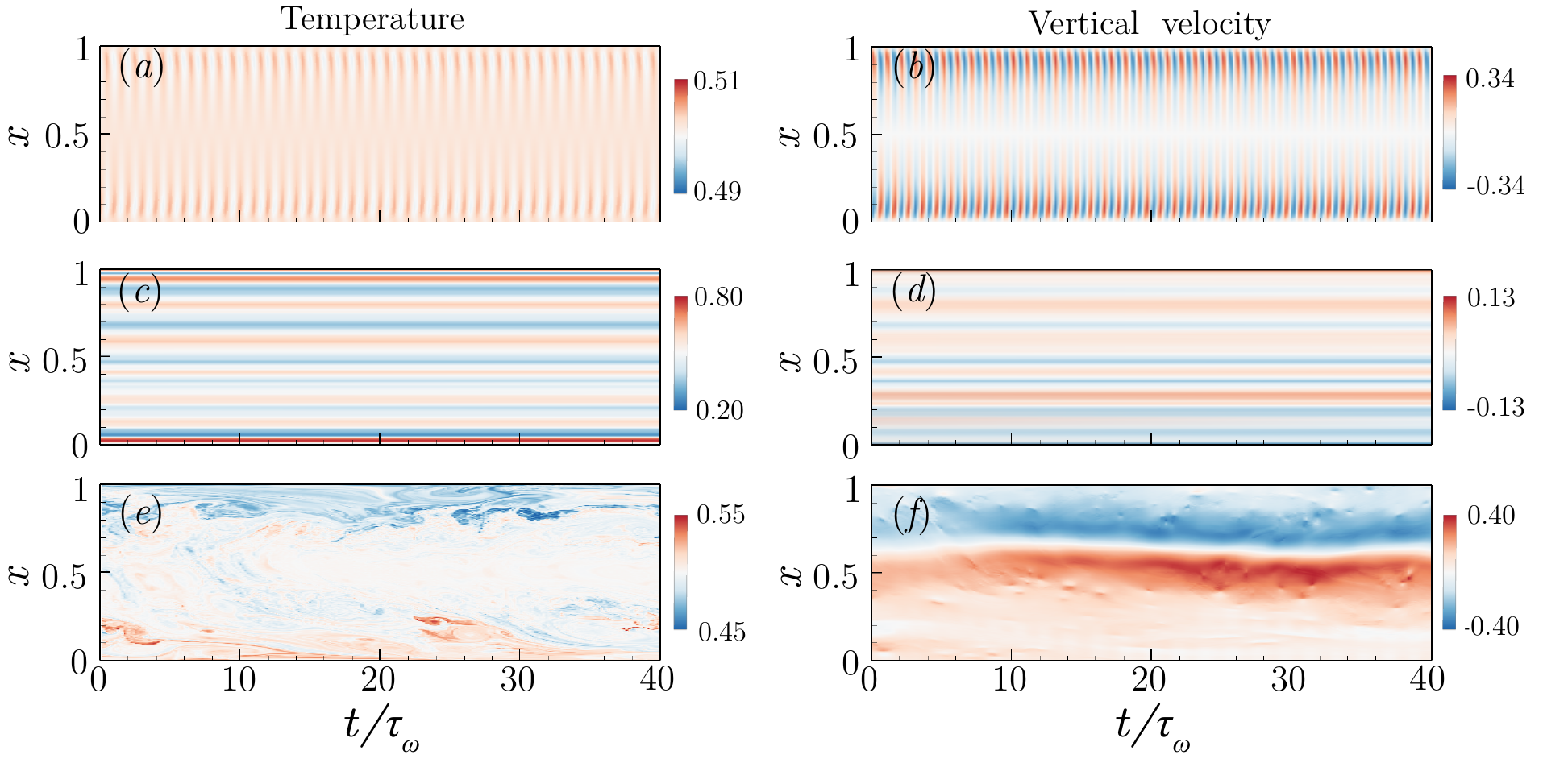}}% Here is how to import EPS art
	\caption{Temporal evolution of temperature $ T(x) \vert_{z=H/2}$  (left column) and vertical velocity $ w(x) \vert_{z=H/2}$ (right column) at the mid-height  for three typical cases: ($a$-$b$) $a=0.01, \omega =10^{3}$ in periodic-circulation regime, ($c$-$d$) $a=0.01, \omega =10^{7}$ in columnar regime, and ($e$-$f$) $a=0.1, \omega =10^{7}$ in columnar-broken regime. Horizontal axis represents the time normalisied by the vibration period $\tau_{\omega}$. }
	\label{fig:profileContour} 
\end{figure}

To investigate the behavior of velocity fluctuations, we examine the time evolution of temperature and velocity profiles at mid-height over forty vibration periods, as shown in figure~\ref{fig:profileContour}. We consider three representative cases corresponding to different flow structures: periodic oscillating circulation, columnar structure, and broken-columnar structure. For the cases with periodic oscillating circulation (small enough $\omega$ shown in figure~\ref{fig:profileContour}$a$,$b$), the continuous reversal of the circulating flow due to simple harmonic vibration leads to periodic changes in temperature and velocity fluctuations, with the period of change matching the imposed vibrating period. As the columnar structure forms (figure~\ref{fig:profileContour}$c$,$d$), the periodic signature fades, and stable strips in the temporal map appear as a footprint of the stable columnar structure. When the frequency is further increased, reaching the regime with broken thermal columns, the strips vanish, and fragmented plumes continuously travel near the sidewall (figure~\ref{fig:profileContour}$e$). However, these temperature anomalies decouple from the region of large vertical velocities, with the maximum velocity occurring mainly in the bulk (figure~\ref{fig:profileContour}$f$). Although the flywheel structure observed here resembles that in RB convection, they are essentially different because the flow structure here contains maximum vertical velocity in the bulk, rather than near the sidewalls.

\subsection {Local velocity standard deviations near the boundaries and at mid-height} %4.3

\begin{figure}
	\centerline{\includegraphics[width=1.0\textwidth]{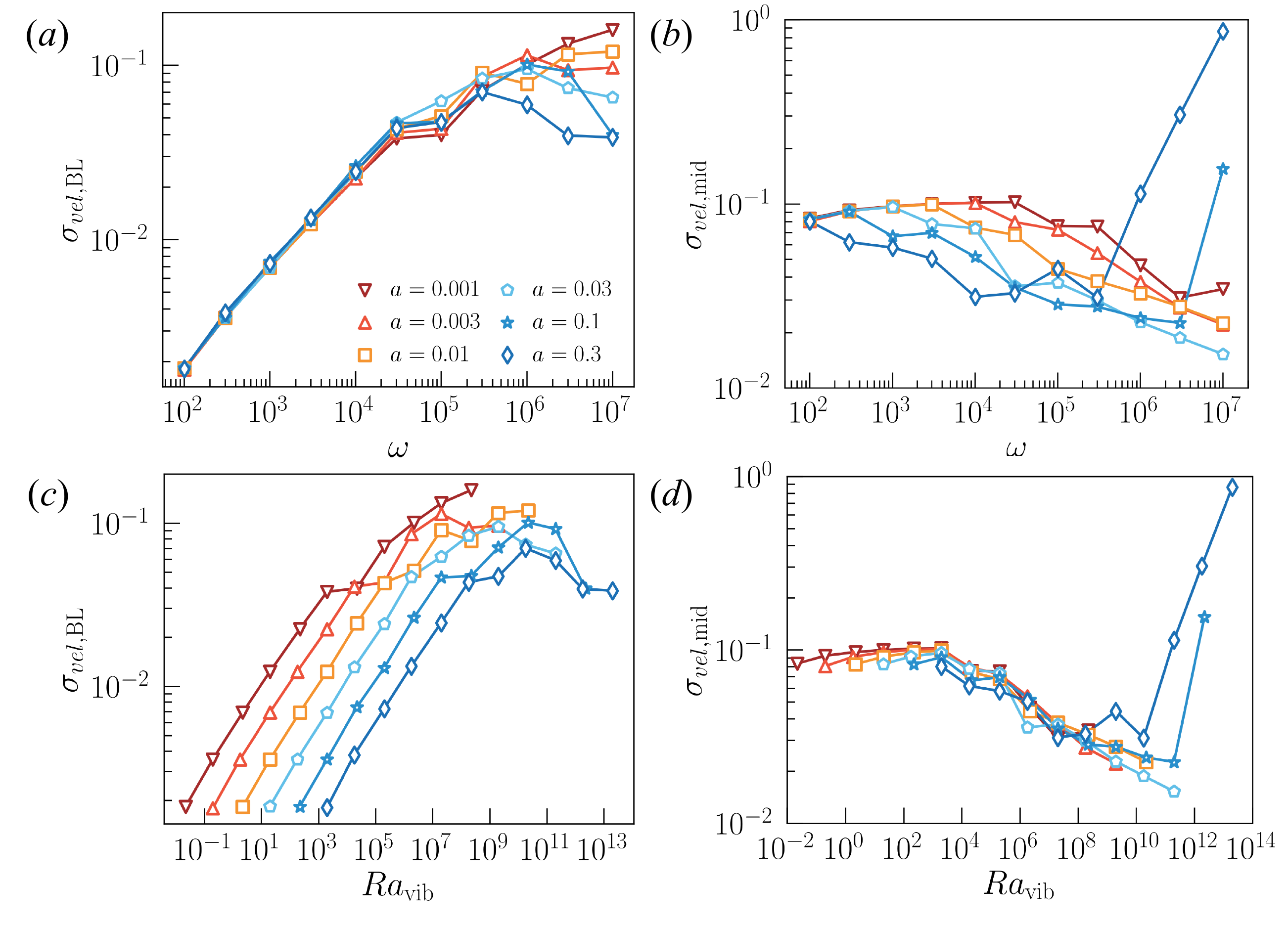}}% Here is how to import EPS art
	\caption{($a$,$c$) Mean velocity SD $\sigma_{\mathrm{vel},\text{BL}}$ at the BL edge, and $(b,d)$ $ \sigma_{\mathrm{vel},\text{mid}}$ at the mid-height as a function of vibration frequency $\omega$ ($a$-$b$) and vibrational Rayleigh number $\Ra_\mathrm{vib}$ ($c$-$d$).}
	\label{fig:GobalaAndMid}
\end{figure}

We further examine the local velocity fluctuations at both the edge of BL and mid-height. The local velocity SD at the BL edge is evaluated as:
\begin{equation}
	\sigma_{\mathrm{vel},\text{BL}}  = \sqrt{\sigma_{u}^{2} \vert_{z=\delta_{\omega}} +  \sigma_{w}^{2}\vert_{z=\delta_{\omega}}  } ,
\end{equation}
and the one at mid-height as:
 \begin{equation}
 	\sigma_{\mathrm{vel},\text{mid}}  = \sqrt{ \sigma_{u}^{2}\vert_{z=H/2}   +  \sigma_{w}^{2}\vert_{z=H/2} } ,
 \end{equation}
respectively. Besides the frequency $\omega$, we introduce $Ra_\mathrm{vib}$, which describes the averaged vibrational effects on convective flows and can be expressed as $Ra_\mathrm{vib} = a^2 \omega^2 Pr /2$, according to~\eqref{eq:ravib}. Our results show that the velocity fluctuations at the BL edge are primarily influenced by the vibration frequency rather than $Ra_\mathrm{vib}$, as evidenced by the collapse of $\sigma_{\mathrm{vel},\text{BL}}$ at small $\omega$ values (figure \ref{fig:GobalaAndMid}$a$-$b$). At positions close to the vibrating plate, the disturbance generated by the plate oscillation propagates as a transverse wave, with a penetration depth given by $\delta_{\omega}/H=\sqrt{2/\omega}$, which only depends on $\omega$. Therefore, the velocity fluctuations at small $\omega$ values exhibit a universal property across different amplitudes. As $\omega$ increases further, the slope of the lines becomes smaller and the curves trend to saturate at high $\omega$. This is probably caused by the fact that at high $\omega$, vibration destabilises the Stokes layer and the velocity fluctuations within the BL are dominant by turbulent fluctuations. The presence of turbulent fluctuations depending on the vibration amplitude may cause the deviations of the curves at high $\omega$.

The parameter $Ra_\mathrm{vib}$ is more relevant than $\omega$ in describing the local velocity SD at mid-height, $\sigma_{\mathrm{vel},\text{mid}}$. Figure~\ref{fig:GobalaAndMid}($b$) shows that $\omega$ does not collapse the data well, whereas $Ra_\mathrm{vib}$ effectively captures the trend of $\sigma_{\mathrm{vel},\text{mid}}$ for different vibration parameters. The plot of $\sigma_{\mathrm{vel},\text{mid}}$ versus $Ra_\mathrm{vib}$ (figure~\ref{fig:GobalaAndMid}$d$) reveals three distinct flow regimes. In regime I with periodic circulation structure, $\sigma_{\mathrm{vel},\text{mid}}$ is not sensitive to changes in $Ra_\mathrm{vib}$. In regime II with columnar structure, $\sigma_{\mathrm{vel},\text{mid}}$ decreases monotonically with increasing $Ra_\mathrm{vib}$, and the decrease in $\sigma_{\mathrm{vel},\text{mid}}$ indicates that columnar plumes become more robust with increasing vibration intensity, leading to the formation of stable columnar structures. In regime III, $\sigma_{\mathrm{vel},\text{mid}}$ suddenly increases sharply, which is associated with the breakdown of the columnar structure.

%\subsection {Phase diagram of flow structure in vibrational thermal convection} %4.3Evolution of Thermal Plumes
\begin{figure}
	\centerline{\includegraphics[width=1.0\textwidth]{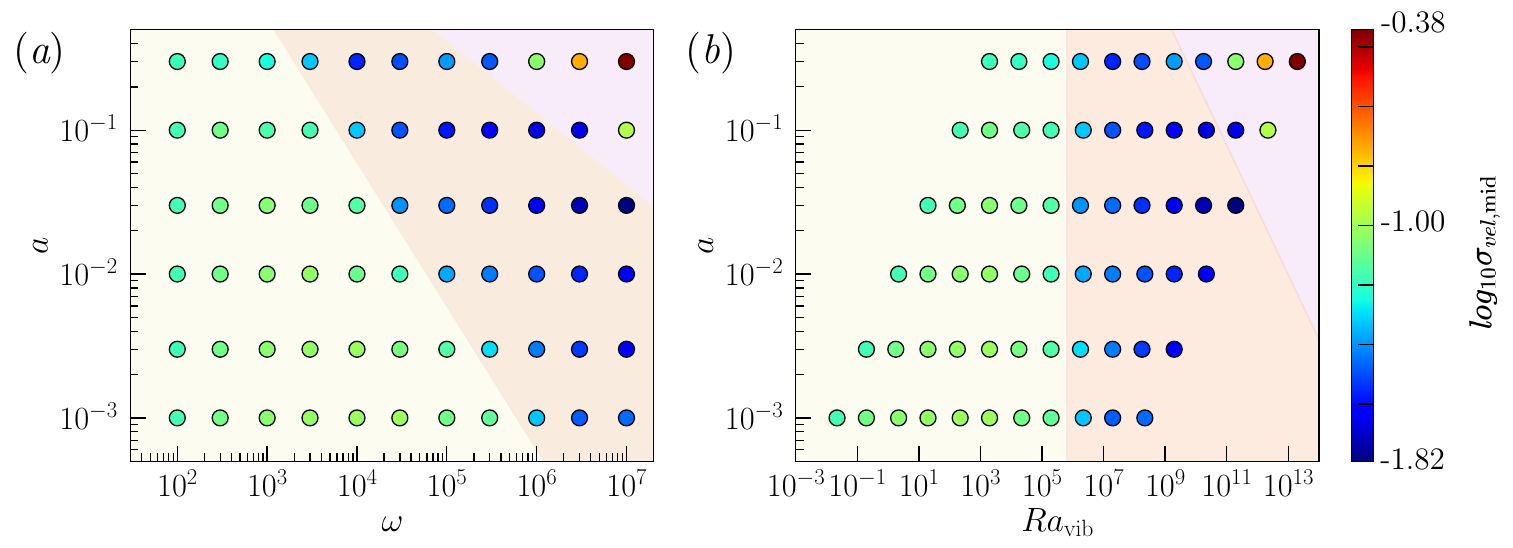}}% Here is how to import EPS art
	\caption{Phase diagram in the $(a,\omega)$ plane ($a$) and in the $(a, \Ra_\mathrm{vib})$ plane ($b$). The color of the points denotes the value of $ log_{10} \sigma_{\mathrm{vel},\text{mid}}$. Three different regimes are represented by different background colours: the periodic-circulation regime (Regime I) coloured by yellow, the columnar regime (Regime II)  coloured by orange, and the columnar-broken regime (Regime III) coloured by purple.}
	\label{fig:phase} 
\end{figure}
Finally, we construct the phase diagram over the full parameter space ($a$,$\omega$) for fixed $Pr = 4.38$. We use $log_{10}\sigma_{\mathrm{vel},\text{mid}}$ to determine the flow regime transitions and the color of the symbol to represent the value of $log_{10}\sigma_{\mathrm{vel},\text{mid}}$. Based on the trend of $log_{10}\sigma_{\mathrm{vel},\text{mid}}$, we can clearly distinguish three regimes: the periodic-circulation regime (yellow-shaded area), the columnar regime (orange-shaded area), and the columnar-broken regime (purple-shaded area) (see figure~\ref{fig:phase}$a,b$). We also examine the transition of the flow regime within the parameter framework of ($a$,$Ra_\mathrm{vib}$). As shown in figure~\ref{fig:phase}$(b)$, $Ra_\mathrm{vib}$ provides a good quantitative description of the transition from regime I to regime II, with a critical vibration Rayleigh number of approximately $Ra_\mathrm{vib} \approx 2.19 \times 10^{6}$. However, the transition from regime II to regime III, that is, from columnar convective structure to LSC, cannot be fully described by $Ra_\mathrm{vib}$.

%{\color{red}From the phase diagram in Fig.~\ref{fig:phase}, We define the regime criteria by the value of $\langle \pmb{u}_{std} \rangle_{V}$.}

\section{\label{sec:level7}Conclusion}\label{sec:conclusion}
In summary, we employed direct numerical simulations to investigate the flow transition in a two-dimensional thermal vibrational convection (TVC) system. Our simulations covered a wide range of dimensionless amplitudes ($0.001 \le a \le 0.3$) and frequencies ($10^{3} \le \omega \le 10^{7}$), with a fixed Prandtl number ($Pr = 4.38$). Our findings indicate that the TVC system exhibits three distinct flow regimes under the influence of vibration with varying amplitudes and frequencies. Based on the analysis of temperature and velocity standard deviations, we classified these regimes as: the periodic-circulation regime, the columnar regime, and the columnar-broken regime. Each regime features different flow structures and statistical properties. In regime I, periodic circulation reversal occurs with the imposed vibration frequency. In regime II, the columnar plume becomes the dominant flow structure. Vibrations within the oscillating boundary layer cause a strong shear effect, destabilizing the thermal boundary layer and triggering the emission of thermal plumes. These plumes merge and self-organize into columnar updrafts or downdrafts, becoming heat channels between the cold and hot plates. These heat columns are almost stationary over space and time, and they are robust. In regime III, breakdown of the heat columns occurs suddenly, and eventually, flywheel structures dominate. Consequently, temperature and velocity fluctuations increase sharply after this breakdown.

Our study has demonstrated that horizontal vibrations can drive the fluid flow and reveal flow state transitions in microgravity TVC systems. Besides its fundamental significance, understanding the transition of flow structure is also important for manipulating fluid devices with dedicated purposes, such as mixing two fluids with different densities. Further research is needed to fully comprehend the regime transitions, such as studying the effects of vibrations on low and very high $Pr$ values instead of fixing $Pr$ at $4.38$.

\section*{Acknowledgements}
This work was supported by the Natural Science Foundation of China under grant nos. 11988102, 11825204, 92052201, 12032016, 11732010, 12102246, and 91852202, the Shanghai Science and Technology Program under Project No. 20ZR1419800, the Shanghai Pujiang program under grant nos. 21PJ1404400, and China Postdoctoral Science Foundation under grant no. 2020M681259.

\section*{Declaration of interests}
The authors report no conflict of interest.

\appendix

\bibliographystyle{jfm}
\bibliography{jfm}

\end{document}